\documentclass[10pt,aps,pra,amsmath,superscriptaddress,twocolumn]{revtex4-1} 
\usepackage{graphicx}
\usepackage{times}
\usepackage{braket}
\usepackage{bm}
\usepackage{color}
\usepackage[caption=false]{subfig}

\makeatletter

\newcommand{\Rmnum}[1]{\expandafter\@slowromancap\romannumeral #1@}
\makeatother

\usepackage[breaklinks]{hyperref}
\hypersetup{colorlinks=true, linkcolor=blue, citecolor=blue, filecolor=blue, urlcolor=blue}

\begin{document}
\title{Nonuniversal entanglement level statistics in
%the area-law phase of 
projection-driven quantum circuits}

\author{Lei Zhang}

\affiliation{Department of Physics, Boston University, Boston, MA 02215, USA}

\author{Justin A. Reyes}

\affiliation{Department of Physics, University of Central Florida, Orlando, FL 32816, USA}

\author{Stefanos Kourtis}

\affiliation{Department of Physics, Boston University, Boston, MA 02215, USA}

\author{Claudio Chamon}

\affiliation{Department of Physics, Boston University, Boston, MA 02215, USA}

\author{Eduardo R. Mucciolo}

\affiliation{Department of Physics, University of Central Florida, Orlando, FL 32816, USA}

\author{Andrei E. Ruckenstein}

\affiliation{Department of Physics, Boston University, Boston, MA 02215, USA}

\begin{abstract}
We study the level-spacing statistics in the entanglement spectrum of
output states of random universal quantum circuits where qubits are
subject to a finite probability of projection to the computational
basis at each time step. We encounter two phase transitions with
increasing projection rate. The first is the volume-to-area law
transition observed in quantum circuits with projective
measurements. We identify a second transition within the area law
  phase by repartioning the system randomly into two subsystems and
  probing the entanglement level statistics. This second transition
  separates a pure Poisson level statistics phase at large projective
  measurement rates from a regime of residual level repulsion in the
  entanglement spectrum, characterized by non-universal level spacing
  statistics that interpolates between the Wigner-Dyson and Poisson
  distributions. By applying a tensor network contraction algorithm
introduced in Ref.~\cite{Yang2017} to the circuit spacetime, we
identify this second projective-measurement-driven transition as a
percolation transition of entangled bonds. The same behavior is
observed in both circuits of random two-qubit unitaries and circuits
of universal gate sets, including the set implemented by Google in its
Sycamore circuits.
\end{abstract}

\maketitle

%%%%%%%%%%%%%%%%%%%%%%%%%%%%%%%%%%%%%%%%%%%%%%%%%%%%%%%%%%%%%%%%%%
\section{Introduction}

Closed quantum many-body systems undergoing unitary evolution
generically reach a thermalized regime, exhibiting volume-law
entanglement~\cite{JDeutsch1991, Calabrese2005, Rigol2008, Kim2013,
  Mezei2017}. Exceptions to this scenario have drawn considerable
attention, due to their relevance to experimentally controllable
quantum systems. For example, many-body localization, which precludes
thermalization and leads instead to area-law entanglement, has been
the subject of extensive theoretical and experimental
work~\cite{Altshuler1997, Huse2007, Huse2010, Bardarson2012,
  Grover2014, PhysRevLett.114.083002, schreiber2015observation,
  choi2016exploring, PhysRevLett.120.070501, smith2016many}. Recently,
failure to thermalize has also been reported in simulations of quantum
circuits subjected to random measurement events that model coupling to
a classical environment~\cite{Cao2018, Chan2018, Skinner2018, Li2019,
  Li2018, Bao2019, Jian2019, Gullans2019}. Interest in these models is
fueled by the ongoing efforts to exploit noisy intermediate-scale
quantum devices for tasks beyond the reach of classical computers.

The studies of Refs.~\cite{Cao2018, Chan2018, Skinner2018, Li2019,
  Li2018, Bao2019, Jian2019, Gullans2019} follow the time evolution of
an initial product state of qubits arranged in a one-dimensional chain
induced by local unitary gates randomly chosen from a volume-law
entangling set (such as, e.g., the Clifford set), and subsequently
measurement of each qubit with probability $p$. Intuitively, the
non-unitary projective measurement operation effectively disentangle
the state and, at sufficiently large measurement rate, results in a
localization of the system in Hilbert space characterized by the
volume-to-area law transition described in Refs.~\cite{Cao2018,
  Chan2018, Skinner2018, Li2019, Li2018, Bao2019, Jian2019,
  Gullans2019}.

In this work, we study volume-law entangling unitary quantum circuits
subjected to a different kind of disentangling perturbation, namely,
projection operations that forcibly ``reset'' qubits to the
computational basis, randomly inserted at a finite rate throughout the
time evolution of the circuit. Furthermore, we employ the level
spacing statistics of the entanglement spectrum~\cite{Li1997},
referred to hereafter as ``the entanglement spectrum statistics" (ESS),
as a finer measure of thermalization and
entanglement~\cite{Chamon2014,Geraedts2016,ZCY2017}. Our computations
are carried out by adopting the iterative tensor network contraction
method introduced in Ref. ~\cite{Yang2017} to the 1+1D spacetime of
the circuits.

%%%%%%%%%%%%%%%%%%%%%%%%%%%%%%%%%%%%%%%%%%%%%%%%%%%%%%%%%%%%%%%%%%%
\begin{figure}[tb]
\includegraphics[width=0.5\textwidth]{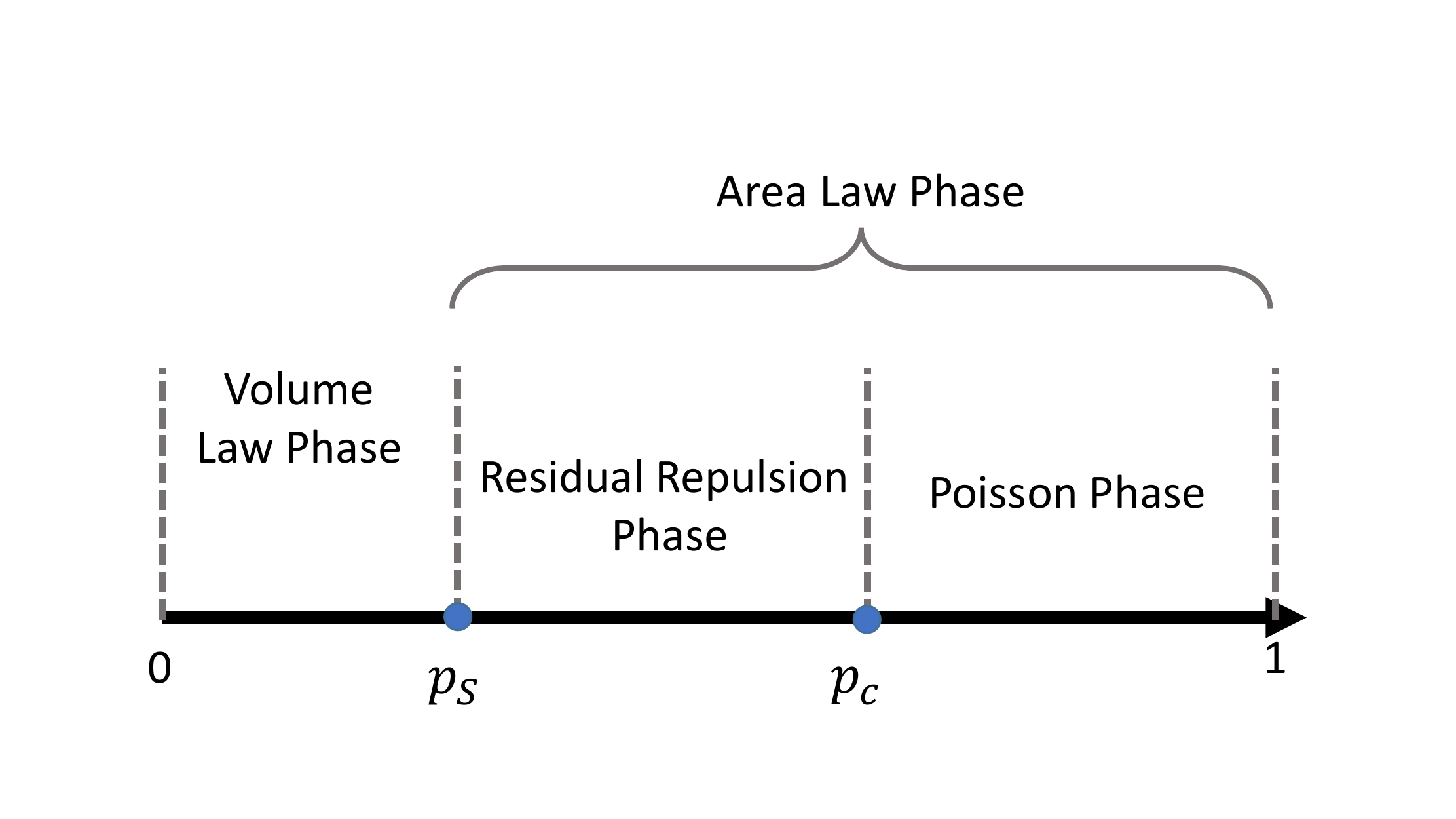}
\caption{Phase diagram as a function of projection probability $p$. We
  observe three phases separated by two phase transitions at $p_S$ and
  $p_c$. The first transition, at $p_S$, is the volume-to-area
  transition identified in Refs.~\cite{Cao2018, Chan2018, Skinner2018,
    Li2019, Li2018, Bao2019, Jian2019, Gullans2019}. The second
  transition separates the pure Poisson level statistics phase
  ($p>p_c$) from a regime of residual level repulsion in the
  entanglement spectrum ($p<p_c$), characterized by non-universal
  level spacing statistics that interpolates between the Wigner-Dyson
  and Poisson distributions. We identify this second
  projective-measurement-driven transition as a percolation transition
  of entangled bonds in 1+1D spacetime.}
\label{fig:illustration}
\end{figure}
%%%%%%%%%%%%%%%%%%%%%%%%%%%%%%%%%%%%%%%%%%%%%%%%%%%%%%%%%%%%%%%%%%%

The main result of this paper is that the EES features three
qualitatively different regimes separated by two transitions. The
first transition displays the same phenomenology as discussed in the
case of random projective measurements, namely, a volume-to-area law
transition at a finite projection rate, $p = p_S$ \cite{foot1}, for
contiguous partitions of the system into subsystems $A$ and
$B$. Within the area law regime, we analyze the EES by changing the
way the system is partitioned, assigning randomly the sites to either subsystem
$A$ or $B$. Probing this repartioned system, we identify a second
transition inside the area law regime for contiguous partitions. For
$p\to {p_S}_+$ the ESS obeys Wigner-Dyson statistics; for $p_S < p <
p_c$, the ESS assumes a non-universal form that interpolates between
the Wigner-Dyson and Poisson statistics (see
Fig.~\ref{fig:illustration}); and for $p>p_c$ the ESS becomes
Poisson-distributed. By resolving the entanglement bond dimensions of
the 1+1D network spatially, we conclude that the transition from
non-universal to Poisson statistics at $p_c$ is associated with the
percolation of entangled bonds in the spacetime geometry of the
circuit. We observe the same behavior, with two transitions, in 1+1D
circuits comprised of either two-qubit random Haar unitaries or of
gates drawn from universal sets, including that implemented by Google
in its Sycamore circuits.

This paper is organized as follows. In Sec.~\ref{sec:circuit}, we
detail the construction of our quantum circuit and of the random
one-qubit projection operators, and outline the method used to compute
the ESS.  We then map each of these circuits into a tensor network and
describe the algorithm for contracting these networks in
Sec.~\ref{sec:tensor}. In Sec.~\ref{sec:results}, we show numerical
results for the ESS in the thermalizing, non-universal, and Poisson
phases and locate the transition point, $p=p_c$, between the latter
two, which we interpret as a two-dimensional percolation transition in
the spacetime of the circuit in
Sec.~\ref{sec:discussion}. Sec.~\ref{sec:conclusion} summarizes our
conclusions.

%%%%%%%%%%%%%%%%%%%%%%%%%%%%%%%%%%%%%%%%%%%%%%%%%%%%%%%%%%%%%%%%%%

\section{Quantum Circuits and Random Matrix Theory}
\label{sec:circuit}

\begin{figure}[tb]
\includegraphics[width=0.5\textwidth]{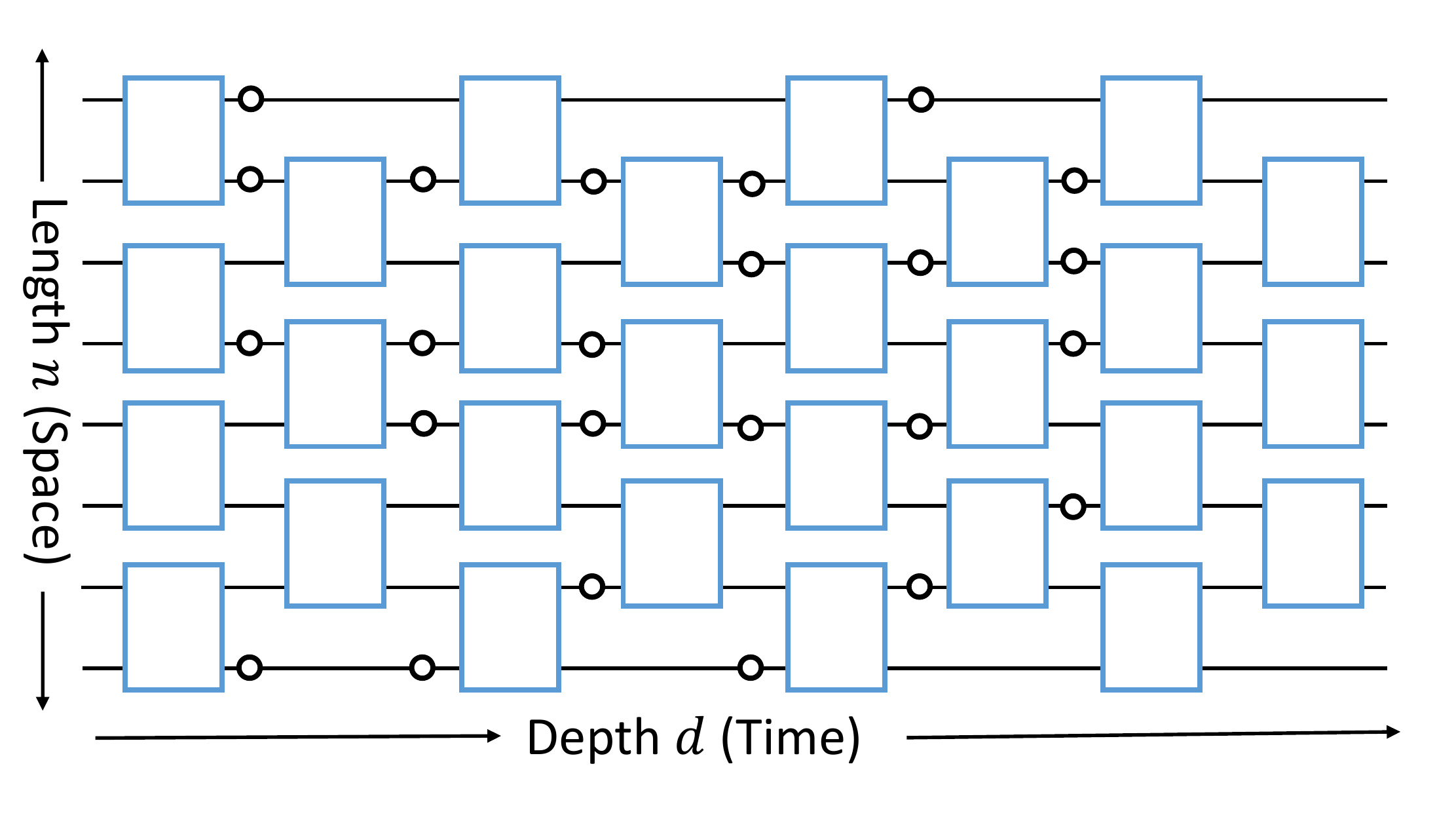}
\caption{Illustration of initial product state evolved in time in a
  quantum circuit. The circuit consists of local two-qubit unitary
  gates (blocks) and projection operators (circles). The latter are
  introduced randomly with probability $p$.}
\label{fig:circuit}
\end{figure}

We consider $n$ qubits evolving in time $t$ from an initial product
state of the form
\begin{equation}
  \label{eq:initial}
  \ket{\Psi(t=0)} = \ket{\psi_1}\otimes \ket{\psi_2}\otimes \dots
  \otimes \ket {\psi_n} \,,
\end{equation}
where the single-qubit state for the $j$-th qubit is defined as
$\ket{\psi_j} = \mathrm{cos}(\theta_j /2)\ket 0 +\mathrm{sin}(\theta_j
/2)e^{i\phi_j}\ket 1$ with arbitrary angles $\theta_j$ and $\phi_j$.
In what follows, the initial state evolves under the action of (i)
random unitary gates, and (ii) single-qubit projection operators,
randomly inserted after each gate with a finite probability $p$. The
state at time $t$ is
\begin{equation}
\ket{\Psi(t)} = M\ket{\Psi(t =0)} = \sum_x \Psi_x(t)\ket x \,,
\end{equation}
where $\ket x =\ket{x_1x_2\dots x_n}$ is a configuration in the
computational basis with $x_j = 0, 1$ for $j = 1,\dots, n$, and $M$ is
a $2^n \times 2^n$ non-unitary matrix describing both the unitary
evolution and the projection operations. The resulting circuit is
illustrated in Fig.~\ref{fig:circuit}, where two-qubit gates are
represented as blocks and projection operators as circles.

We choose the projection operator acting on the $j$-th qubit to take
the form $M_{0} = I_1 \otimes I_2 \otimes\dots \otimes \ket {0_j} \bra
{0 _j} \otimes \dots \otimes I_n$, where $I_j$ is the identity
operator on a single qubit. Although we have chosen to project to the
$\ket 0$ instead of the $\ket 1$ state, this choice is immaterial in
what follows. Projection operators are not norm-preserving, and hence
the final state $\ket{\Psi(t)}$ is not normalized by default. We
normalize final states for consistency. Projection operators can be
physically interpreted as randomly picking a qubit and resetting it to
the computational basis.  As will become evident below, the projector
operators have a disentangling effect, similar to that of the addition
of measurement operators in random Clifford or Haar-random
circuits~\cite{Cao2018, Chan2018, Skinner2018, Li2019, Li2018,
  Bao2019, Jian2019, Gullans2019}. In particular, projectors also lead
to a volume-to-area law transition as a function of nonunitary
operator density.

%%%%%%%%%%%%%%%%%%%%%%%%%%%%%%%%%%%%%%%%%%%%%%%%%%%%%%%%%%%%%%%%
%%%%%%%%%%%%%%%%%%%%%%%%%%%%%%%%%%%%%%%%%%%%%%%%%%%%%%%%%%%%%%%%

Now, consider the pure state $|\Psi\rangle = \sum_{x}\Psi_{x}
|x\rangle$ after evolution with a quantum circuit as specified above,
where $x$ is the configuration of the qubits in the computational
basis and $\Psi_{x}$ is the coefficient for each $x$. $\Psi_{x}$ can
be reshaped into a matrix $\Psi_{A,B}$ by splitting the state into
subsystems $A$ and $B$. In this way, $\ket{\Psi}$ is expressed as
\begin{equation}
|\Psi\rangle = \sum_{x_A,x_B}\Psi_{A,B} \ket{x_A}\otimes \ket{x_B}\,,
\end{equation}
where $x_A$ and $x_B$ are the local configurations for subsystems $A$
and $B$, respectively. Notice that the partition into subsystems
  $A$ and $B$ can be chosen arbitrarily; for example, the subsystems
  can be each contiguous or not. The choice of contiguous subsystems
  is often made to study if the entanglement entropy satisfies volume
  or area law. Here we shall also consider random partitions, in which
  sites are randomly assigned to subsystem $A$ or $B$. This partitioning 
allows us to probe the ESS even when the entanglement entropy is small
for contiguous partition, e.g., in the area law regime.

The entanglement spectrum can be obtained by a
Schmidt decomposition~\cite{Li1997}
\begin{equation}
|\Psi\rangle = \sum_{k} \lambda_k \ket{x_A^k}\otimes \ket{x_B^k}\,,
\end{equation}
which is equivalent to singular value decomposition (SVD) of matrix
$\Psi_{A,B}$ with singular values $\lambda_k$.

The set of entanglement levels $\lambda_k$ defines the entanglement
spectrum (ES). The entanglement entropy is given by
\begin{equation}
S = -\sum_k \lambda_k^2\, \ln \lambda_k^2 \,.
\end{equation}
With the ES in descending order, $\lambda_k > \lambda_{k+1}$, the
ratio of adjacent gaps in the spectrum can be defined as
\begin{equation}
r_k = \frac{ \lambda_{k-1}- \lambda_{k}}{ \lambda_{k}- \lambda_{k+1}} \,.
\end{equation}
We remark that ratios of other functions of the $\lambda_k$ can
  be defined, but these choices do not affect the ESS. For example,
  let us define
\begin{align}
  r^{(f)}_k
  &=
  \frac{f(\lambda_{k-1})-f(\lambda_{k})}{f(\lambda_{k})-f(\lambda_{k+1})}
  \\
  &=
  r_k
  \times
  \frac{f(\lambda_{k-1})-f(\lambda_{k})}{\lambda_{k-1}-\lambda_{k}}
  \Big/
  \frac{f(\lambda_{k})-f(\lambda_{k+1})}{\lambda_{k}-\lambda_{k+1}}
  \; \nonumber,
\end{align}
which in the limit when the level spacings go to zero can be replaced
by
\begin{align}
  r^{(f)}_k
  &\to
  r_k
  \times
  \frac{f'(\lambda_{k-1})}{f'(\lambda_{k})}
  \to
  r_k
  \;.
\end{align}
In other words, for any smooth function $f$ the statistics for the
ratios $r^{(f)}_k$ are the same as those for the ratios $r_k$, in the
limit of dense spectra or small $\lambda_{k-1}-\lambda_k$ spacings.

For Haar-random states, the probability distribution for adjacent
entanglement level ratios, which defines the ESS, follows Wigner-Dyson
statistics from random matrix theory~\cite{Yang2015,Marcenko1967} and
fits well the surmise \cite{Atas2013}
\begin{equation}
P_{\mathrm{WD}}(r) =
\frac{1}{Z}\frac{(r+r^2)^{\beta}}{(1+r+r^2)^{1+3\beta/2}}
\end{equation}
with $Z = 4\pi/81\sqrt{3}$ and $\beta = 2$ for the Gaussian Unitary
Ensemble (GUE) distribution. In contrast, the ESS for integrable
systems takes the Poisson form
\begin{equation}
P_{\mathrm{Poisson}}(r) = \frac{1}{(1+r)^2} \,.
\end{equation}
The most marked difference between the GUE and Poisson distributions
is the level repulsion ($P_{\mathrm{WD}} \rightarrow 0$ for
$r\rightarrow 0$) in the former and its absence ($P_{\mathrm{Poisson}}
> 0$ at $r=0$) in the latter.

%%%%%%%%%%%%%%%%%%%%%%%%%%%%%%%%%%%%%%%%%%%%%%

\section{Quantum Circuits as Tensor Networks}\label{sec:tensor}

%%%%%%%%%%%%%%%%%%%%%%%%%%%%%%%%%%%%%%%%%%%%%%%%%%%%%%%%%%%%%%%%%%%%%%%%%%%%
\subsection{Tensor network mapping}

%%%%%%%%%%%%%%%%%%%%%%%%%%%%%%%%%%%%%%%%%%%%%%%%%%%%%%%%%%%%%%%%%%%%%%%%%%%%
\begin{figure}[tb]
\includegraphics[width=0.5\textwidth]{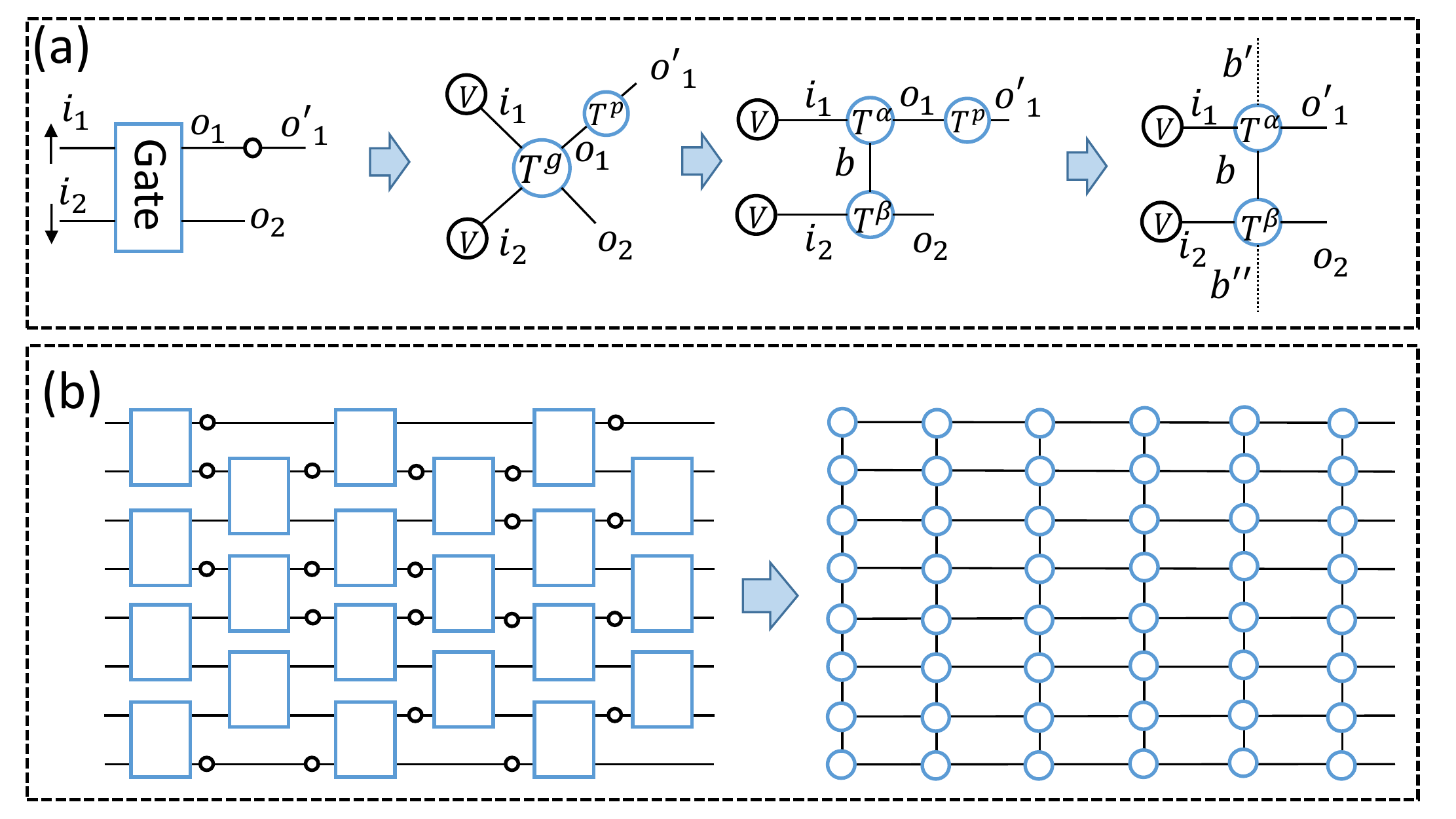}
\caption{Panel (a) shows how to rewrite a local two-qubit Haar random
  gate followed by a projection operator into two tensors where each
  one has four active ($i_1,i_2,o_1,o_2$) and two dummy ($b,b'$)
  indices. The active indices can be lumped together as $i_1i_2$ and
  $o_1o_2$. Panel (b) shows how to use the correspondence between two-
  and one-bit gates and a rank-4 tensor to map a quantum circuit into
  a rectangular tensor network.}
\label{fig:mapping}
\end{figure}
%%%%%%%%%%%%%%%%%%%%%%%%%%%%%%%%%%%%%%%%%%%%%%%%%%%%%%%%%%%%%%%%%%%%%%%%%%%%

In this section, we map the random quantum circuits introduced above
and illustrated in Fig.~\ref{fig:illustration} into a square-grid
tensor network and detail the contraction algorithm we use to compute
entanglement properties. The tensorial representation of the elements
of the circuits is illustrated in Fig.~\ref{fig:circuit}. We use
  open boundary conditions in the space and time directions.

Each two-qubit gate $g$ is expressed as a $4\times 4$ unitary matrix
$T^{g}_{(i_1i_2)(o_1o_2)}$, where $(i_1i_2)$ and $(o_1o_2)$ are
combined indices corresponding to gate input and output qubit states,
respectively. Each $4\times 4$ matrix can be reshaped into a
$2\times2\times2\times2$ tensor $T^{g}_{i_1i_2o_1o_2}$.  Since we want
to transform the circuit into a square lattice geometry, we regroup
the indices of each tensor to reshape it to a matrix
$T^{g}_{(i_1o_1)(i_2o_2)}$ and use a SVD to decompose it as
\begin{align}
 T^{g}_{(i_1o_1)(i_2o_2)} =&{\ } \sum_{m, m'} U_{(i_1o_1)m}\,
 \Sigma_{mm'}\, V^{\star}_{(i_2o_2)m'} \nonumber \\ = &{\ } \sum_{m,
   b,m'} U_{(i_1o_1)m}\, \sqrt{\Lambda}_{mb}\, \sqrt{\Lambda}_{bm'}\,
 V^{\star}_{(i_2o_2)m'} \nonumber \\ =&{\ } \sum_{b}
 T^{\alpha}_{(i_1o_1)b}\, T^{\beta}_{(i_2o_2)b} \,,
\end{align}
where $U_{(i_1o_1)m}$ and $V_{(i_2o_2)m'}$ are unitary matrices and
$\Lambda_{mm'}$ is a semi-positive diagonal matrix containing the
singular values. The two new matrices $ T^{\alpha}_{(i_1o_1)b}
=\sum_{m} U_{(i_1o_1)m}\sqrt{\Lambda}_{mb} $ and $
T^{\beta}_{(i_2o_2)b} = \sum_{m'}\sqrt{\Lambda}_{bm'}
V^{\star}_{(i_2o_2)m'} $, with $b$ an index running over singular
values, are then reshaped to tensors $ T^{\alpha}_{i_1o_1b} $ and $
T^{\beta}_{i_2o_2b} $. To end up with a rectangular geometry, we add a
fourth ``dummy'' index with dimension 1 to each tensor, connecting it
with a neighboring tensor in the space dimension, as indicated by the
faint vertical lines in Fig.~\ref{fig:mapping}a.

Each single-qubit projector can also be expressed as a tensor
$T^{p}_{o_1 o_1'}$, where $o_1'$ has dimension 1. These can be
contracted into gate tensors as
\begin{equation}
 T^{\alpha}_{i_1o_1'b b'} = \sum_{o_1} T^{\alpha}_{i_1o_1b b'}\,
 T^{p}_{o_1 o_1'} \,.
\end{equation}
Since initial states are taken to be product states, they can be
written simply as a tensor product of vectors, each vector
corresponding to a single-qubit state. The state for the first qubit,
for example, is $\ket{\psi_{i_1}} = [\mathrm{cos}(\theta_{i_1}/2),
  \mathrm{sin}(\theta_{i_1}/2) e^{\phi_{i_1}}] = V_{i_1}$. This can be
contracted into the first gate tensor as
\begin{equation}
  T^{\alpha}_{o_1'b b'} = \sum_{i_1} V_{i_1}\, T^{\alpha}_{i_1o_1'b
    b'} \,.
\end{equation}
Finally, wherever no gates are applied to qubits at the top and bottom
boundaries, a rank-3 identity tensor $\delta_{i_1 o_1 b}$ is added to
complete the square lattice. With the above transformations, we map
the evolution described by the quantum circuit into a tensor network,
as shown in Fig.~\ref{fig:mapping}b. Note that the final (right)
column of $n$ indices is left free.

Next, we adopt a common indexing scheme for all tensors in the
network, where we denote every single tensor index as $s$. The set of
all indices in the tensor network is thus $\{\textbf{S}\} =
\{s_1,s_2,...,s_N\}$, where $N = 2d(2n-1)$ is the total number of
indices and $d$ is the circuit depth in time steps with 2 columns of
two-qubit gates per time step. Each tensor can be uniquely determined
by its subset of indices $\{\textbf{s}\} \subset \{\textbf{S}\}$ as
$T_{\{\textbf{s}\}}$. In this language, the final state $\ket{\Psi_f}$
is
\begin{equation}
\ket{\Psi_f} =  \mathrm{Tr} \prod_{\{\textbf{s}\}}T_{\{\textbf{s}\}},
\end{equation}
where Tr indicates a trace over all non-free indices connecting
tensors. Obtaining the final state is thus equivalent to partially
contracting a tensor network.

%%%%%%%%%%%%%%%%%%%%%%%%%%%%%%%%%%%%%%%%%%%%%%%%%%%%%%%%%%%%%%%%%%%%%%%%%%%%
\subsection{Contraction algorithm}

%%%%%%%%%%%%%%%%%%%%%%%%%%%%%%%%%%%%%%%%%%%%%%%%%%%%%%%%%%%%%%%%%%%%%%%%%%%%
\begin{figure}[tb]
\includegraphics[width=0.5\textwidth]{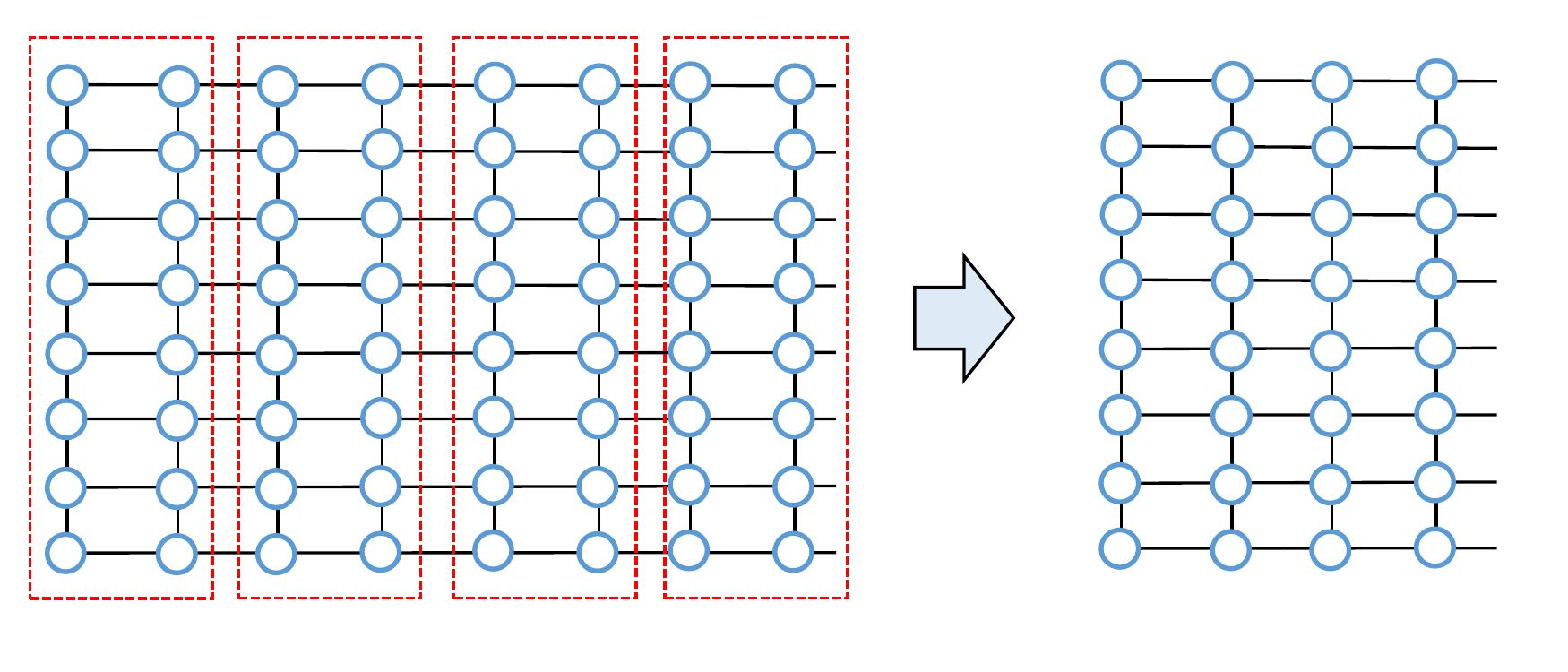}
\caption{Illustration of the coarse-graining process used to contract
  tensor networks in the time direction.}
\label{fig:contraction}
\end{figure}
%%%%%%%%%%%%%%%%%%%%%%%%%%%%%%%%%%%%%%%%%%%%%%%%%%%%%%%%%%%%%%%%%%%%%%%%%%%%

To contract tensor networks, we employ a variant of the iterative
compression-decimation (ICD) algorithm introduced in
Ref.~\cite{Yang2017}. We perform iterations of alternating
\textit{compression} and \textit{decimation} steps until the width of
the lattice is fully contracted.

The compression step is a sweep over lattice bonds where we first
contract the tensors at the ends of each visited bond and then perform
a SVD to restore the structure of the lattice, in a way reminiscent of
the density-matrix renormalization group
algorithm~\cite{Schollwock2011}. This step becomes significant when
the projector density in quantum circuits is increased, as we will
discuss below. We use the tools developed in Ref.~\cite{Yang2017} to
implement compression efficiently.
%We stress that all contraction and
%decomposition steps are exact as only singular values that are
%\textit{precisely} equal to zero (to machine accuracy) are
%discarded.

The decimation step coarsens the lattice at the expense of increasing
bond dimensions. As illustrated in Fig.~\ref{fig:contraction} the
tensor network coarse-graining is performed in time direction, so that
every two columns of tensors are contracted into one. At the end of
decimation, we get a tensor chain representing the final state.

%%%%%%%%%%%%%%%%%%%%%%%%%%%%%%%%%%%%%%%%%%%%%%%%%%%%%%%

%%%%%%%%%%%%%%%%%%%%%%%%%%%%%%%%%%%%%%%%%%%%%%%%%%%%%%%%%%%%%%%%%%%%%%%%%%
\begin{figure*}[t]
\subfloat[][]{
\includegraphics[width=.33\textwidth]{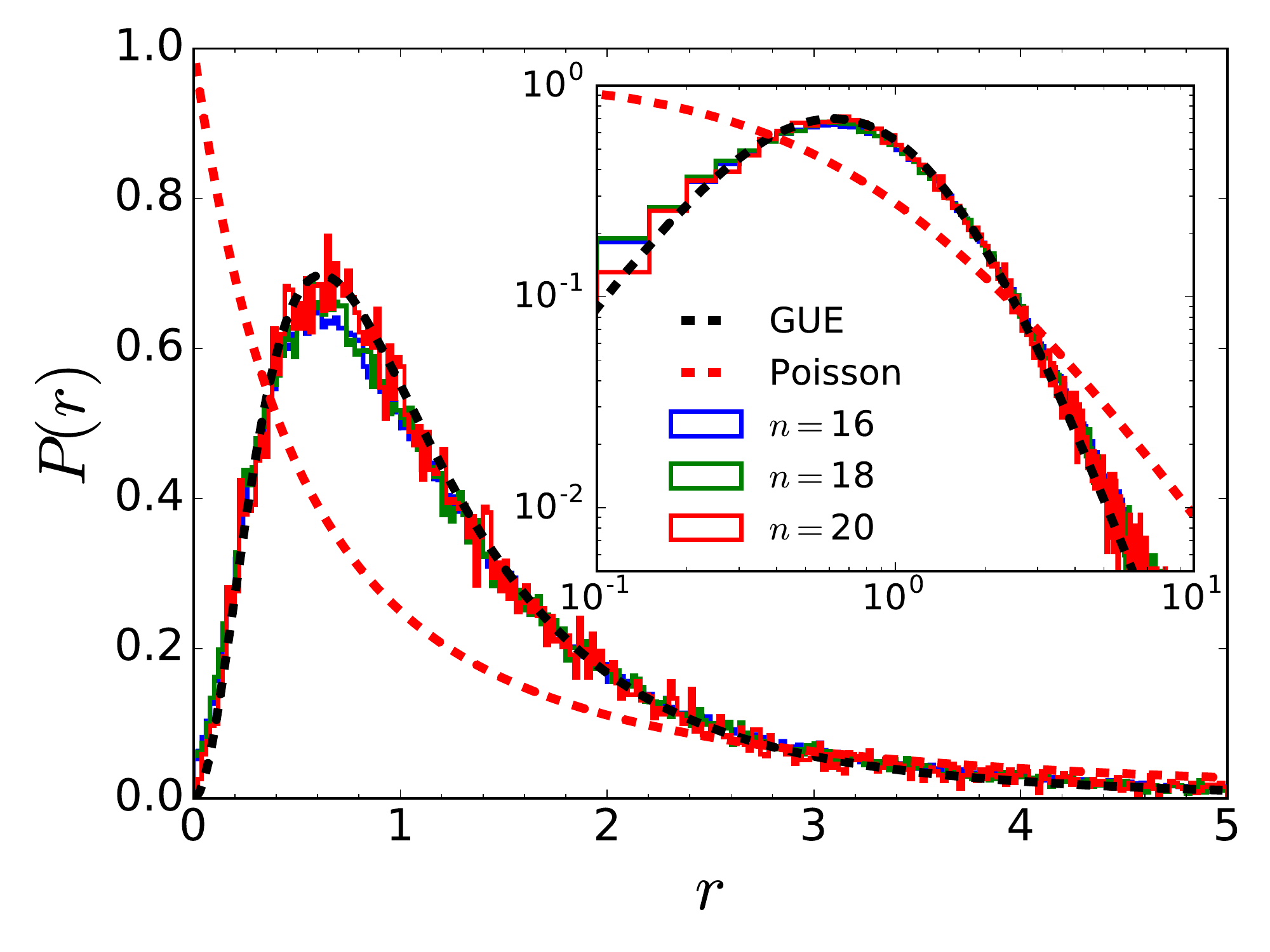}
}
\subfloat[][]{
\includegraphics[width=.33\textwidth]{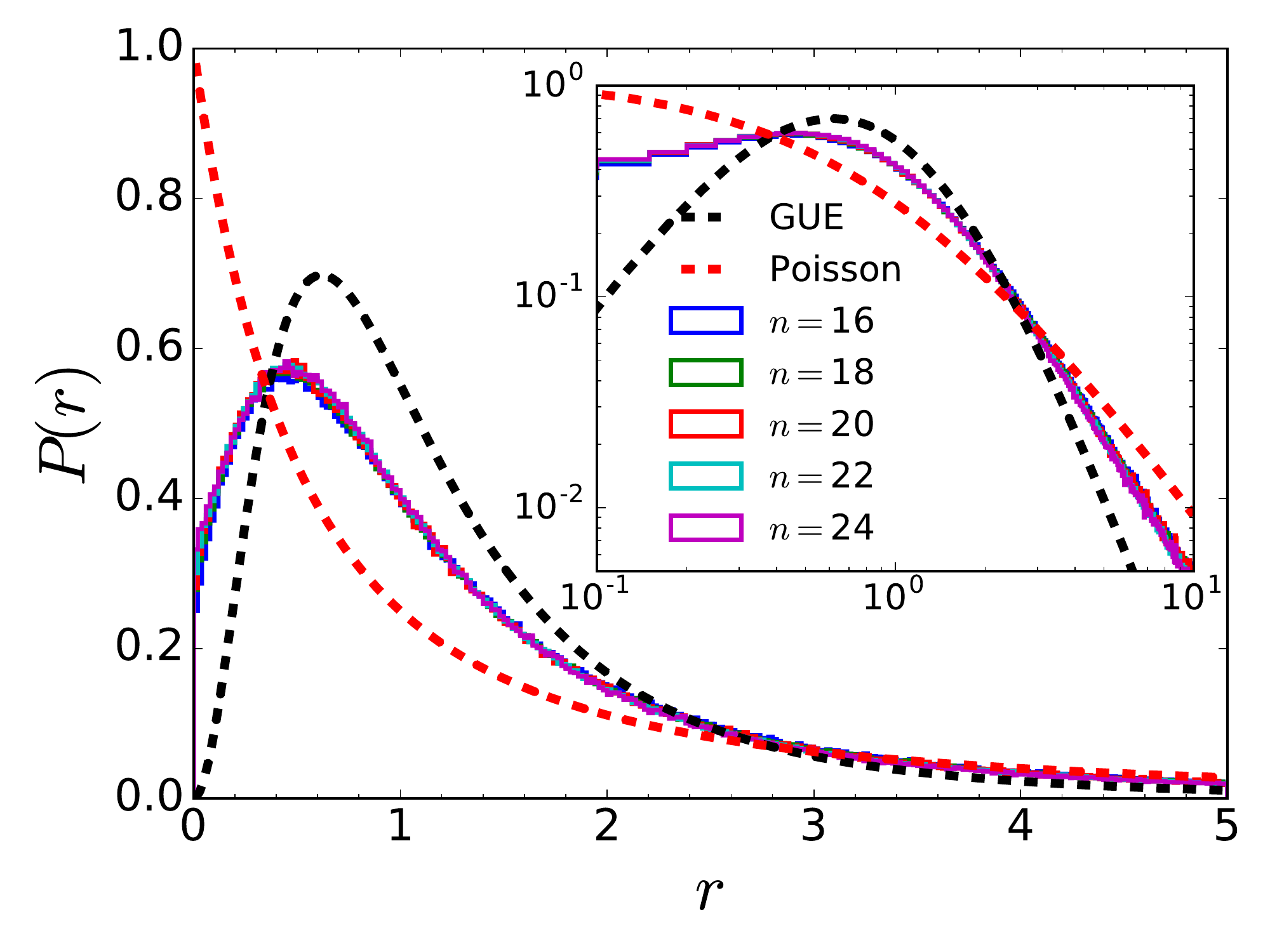}
}
\subfloat[][]{
\includegraphics[width=.33\textwidth]{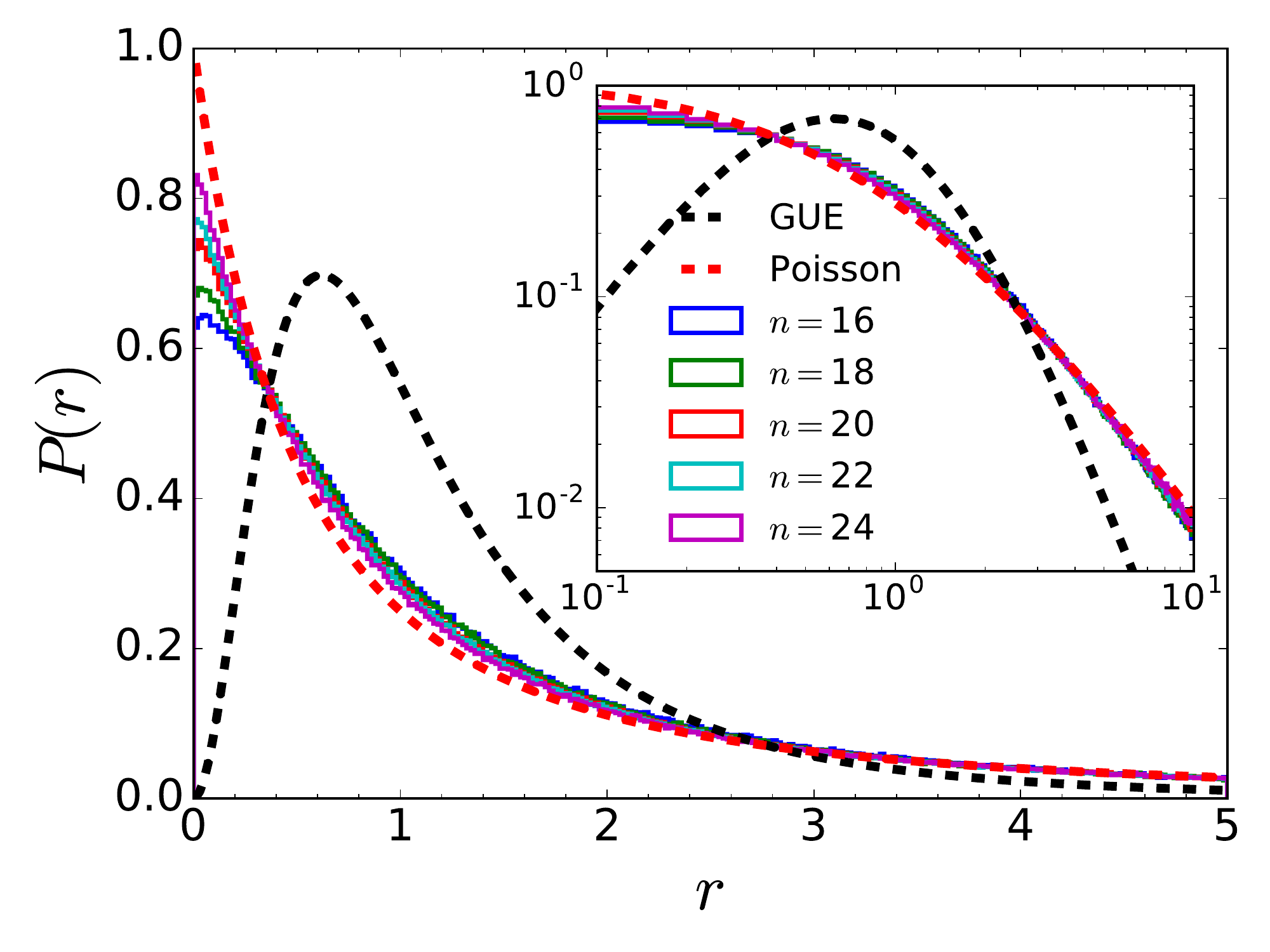}
}\\
\subfloat[][]{
\includegraphics[width=.33\textwidth]{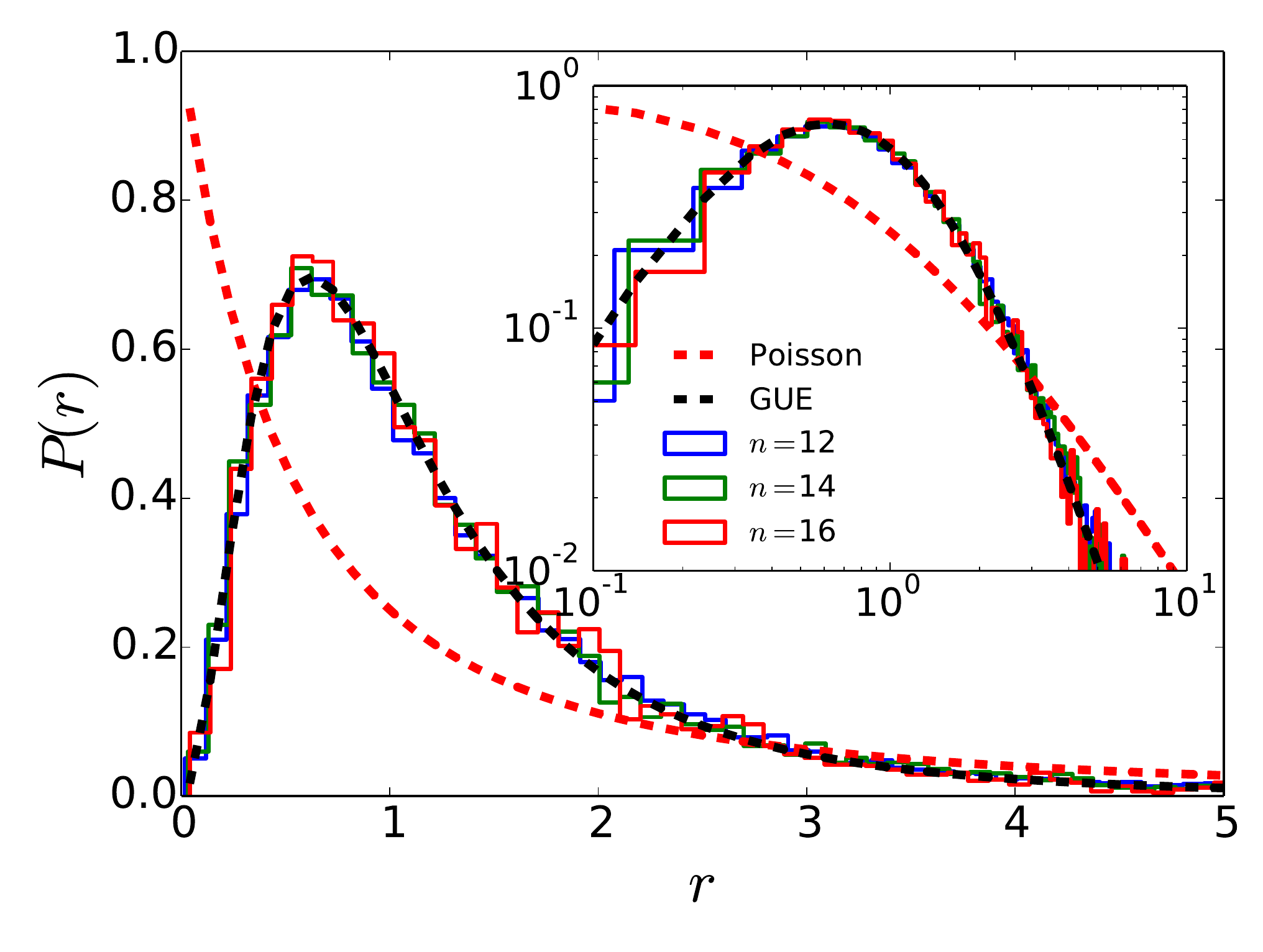}
}
\subfloat[][]{
\includegraphics[width=.33\textwidth]{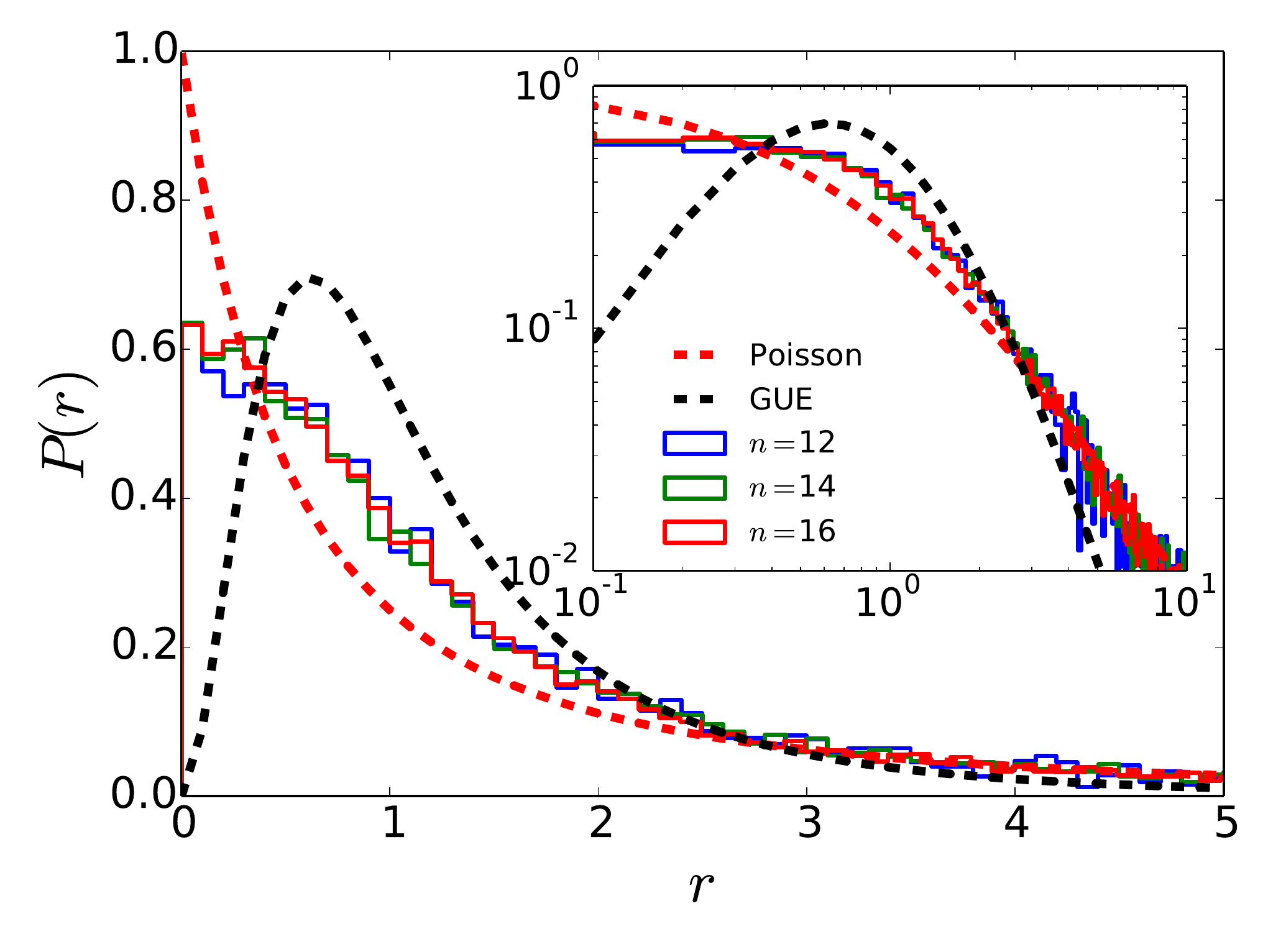}
}
\subfloat[][]{
\includegraphics[width=.33\textwidth]{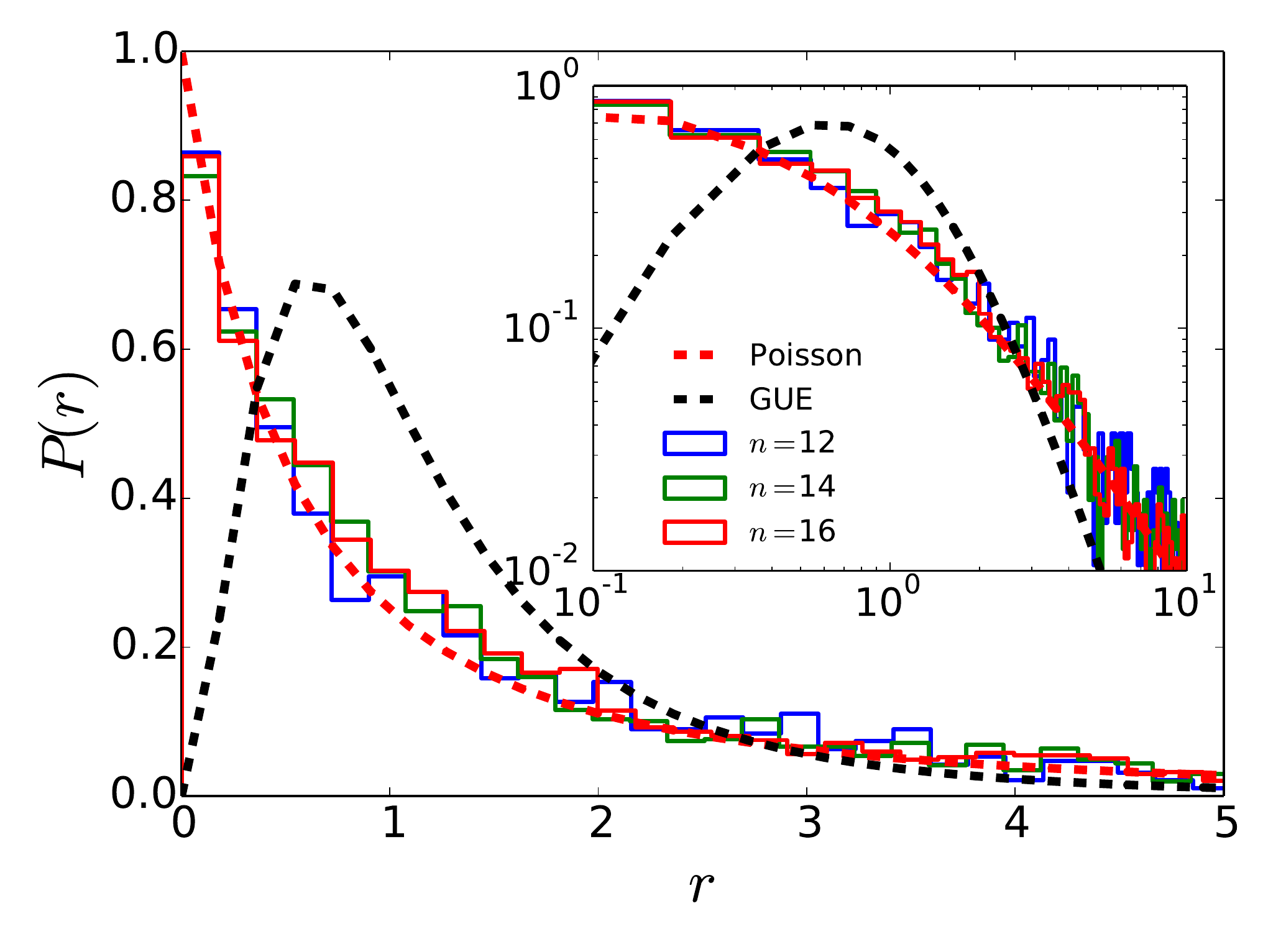}
}
\caption{Level spacing ratio distributions for the entanglement
  spectrum in the three phases of Fig.~\ref{fig:illustration}. From
  (a) to (c), the distributions are taken from circuits constructed
  from the random Haar-measure with $N$ bits and with depth $d=N$,
  whereas from (d) to (f) the distributions are taken from circuits
  constructed from the universal gate set $\mathbf{U}_{\rm{syc}} =
  \sqrt{X},\sqrt{Y},\sqrt{W},\rm{fSim}$ with $N$ bits and depth
  $d=2N$. (a) Volume-law phase at $p=0.2$, as indicated by
  the GUE distribution with level repulsion in the limits $r
  \rightarrow 0$ and a Gaussian tail at $r \rightarrow \infty$. (b)
  Residual repulsion phase at $p=0.35$. Level repulsion disappears at
  $r \rightarrow 0$ while a majority of levels show level repulsion as
  indicated by the presence of a peak at finite $r$. (c) Poisson phase
  at $p = 0.5$. The spectrum displays an absence of level repulsion in
  similarity to the Poisson distribution. (d) Volume-law phase at
  $p=0$, again indicated by the GUE statistics in the level
  spacing. (e) Residual repulsion phase at $p=0.2$ exhibiting a shoulder
 instead of the shifted peak seen in (b). (f) Poisson
  phase at $p=0.4 $. The insets show the distributions in log-log
  scales in order to capture their behavior at the tails. In (a),
  results are obtained from 500 realizations for $n =20$ and from 1000
  reliazations for $n <20$.  For (b) and (c), , results are obtained
  from 1000 realizations for up to $n =24$.  In (d), (e), and (f)
  results are obtained from 500 realizations up to $n=16$.}
\label{fig:spectrum}
\end{figure*}
%%%%%%%%%%%%%%%%%%%%%%%%%%%%%%%%%%%%%%%%%%%%%%%%%%%%%%%%%%%%%%%%%%%%%%%%%%

%%%%%%%%%%%%%%%%%%%%%%%%%%%%%%%%%%%%%%%%%%%%%%%%%%%%%%%%%%%%%%%%%%%%%%%%%%
\section{Entanglement spectrum statistics and phase transitions}
\label{sec:results}

We now use the formulations and tools previously discussed to perform
numerical investigations of the ESS as a function of projector
density, $p$, in random quantum circuits satisfying the geometry given
in Fig.~\ref{fig:circuit}. For each circuit realization, the initial
state is of the form of Eq.~\eqref{eq:initial}, where all $\theta_j$
and $\phi_j$ are selected uniformly at random. For comparative
purposes, two separate gate sets were used to construct the random
circuits. The first case is composed of two-qubit gates selected from
the Haar-random measure, while the second consists of gates uniformly
selected from the universal gate set $\mathbf{U}_{\rm{syc}}=\lbrace
\sqrt{X},\sqrt{Y},\sqrt{W},\rm{fSim}\rbrace$ ~\cite{Arute2019}. A
single-qubit projector is applied with probability $p$ to each qubit
after every gate and before the final time step. We calculate the ES
of many random realizations and bin the spectra for the same $p$ to
obtain the ESS. Our results are summarized in Fig.~\ref{fig:spectrum}.

We begin by characterizing the quantum chaotic and integrable regimes
for these circuits with Wigner-Dyson and Poisson distribution ESS,
respectively, and then proceed to describe a previously unforeseen
intermediate regime. For $p < p_S$, the ESS follows the GUE
distribution over the entire range of $r$, as seen in
Fig.~\ref{fig:spectrum}a,d. This corresponds to a highly entangled
final state (i.e. volume-law state), indicating that the system has
settled into the quantum chaotic regime. On the other hand, for $p >
p_c$, with $p_c \simeq 0.41$, the ESS follows the Poisson
distribution, which indicates that the system is integrable --- see
Fig.~\ref{fig:spectrum}c,f. The intuition for this change in the ESS
as a function of $p$ is that as the frequency of projectors is
increased, the system becomes frozen in local states, therefore
failing to entangle~\cite{Cao2018, Chan2018, Skinner2018, Li2019,
  Li2018, Bao2019, Jian2019, Gullans2019}. The small deviations from
the expected exact Poisson distribution are due to our choice to avoid
placing projectors in the final time step of simulations. This choice
prevents us from potentially reducing the number of qubits in the
system at the final time step. This effect disappears in the
thermodynamic limit, which is investigated below by analysis using finite size scaling.

%%%%%%%%%%%%%%%%%%%%%%%%%%%%%%%%%%%%%%%%%%%%%%%%%%%%%%%%%%%%%%%%%%%%%%%%%%
\begin{figure*}[t]
\subfloat[][]{
\includegraphics[width=.33\textwidth]{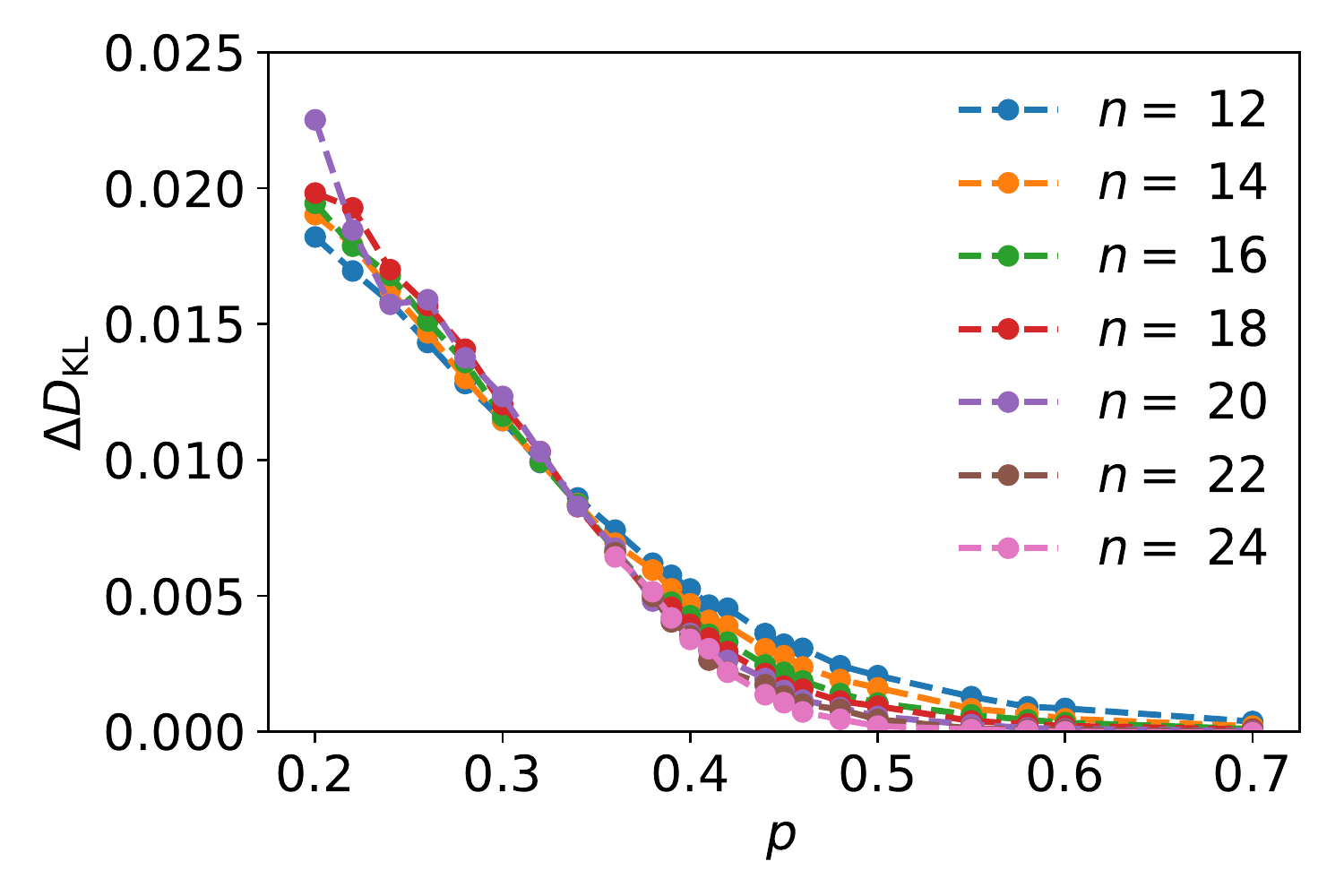}
}
\subfloat[][]{
\includegraphics[width=.33\textwidth]{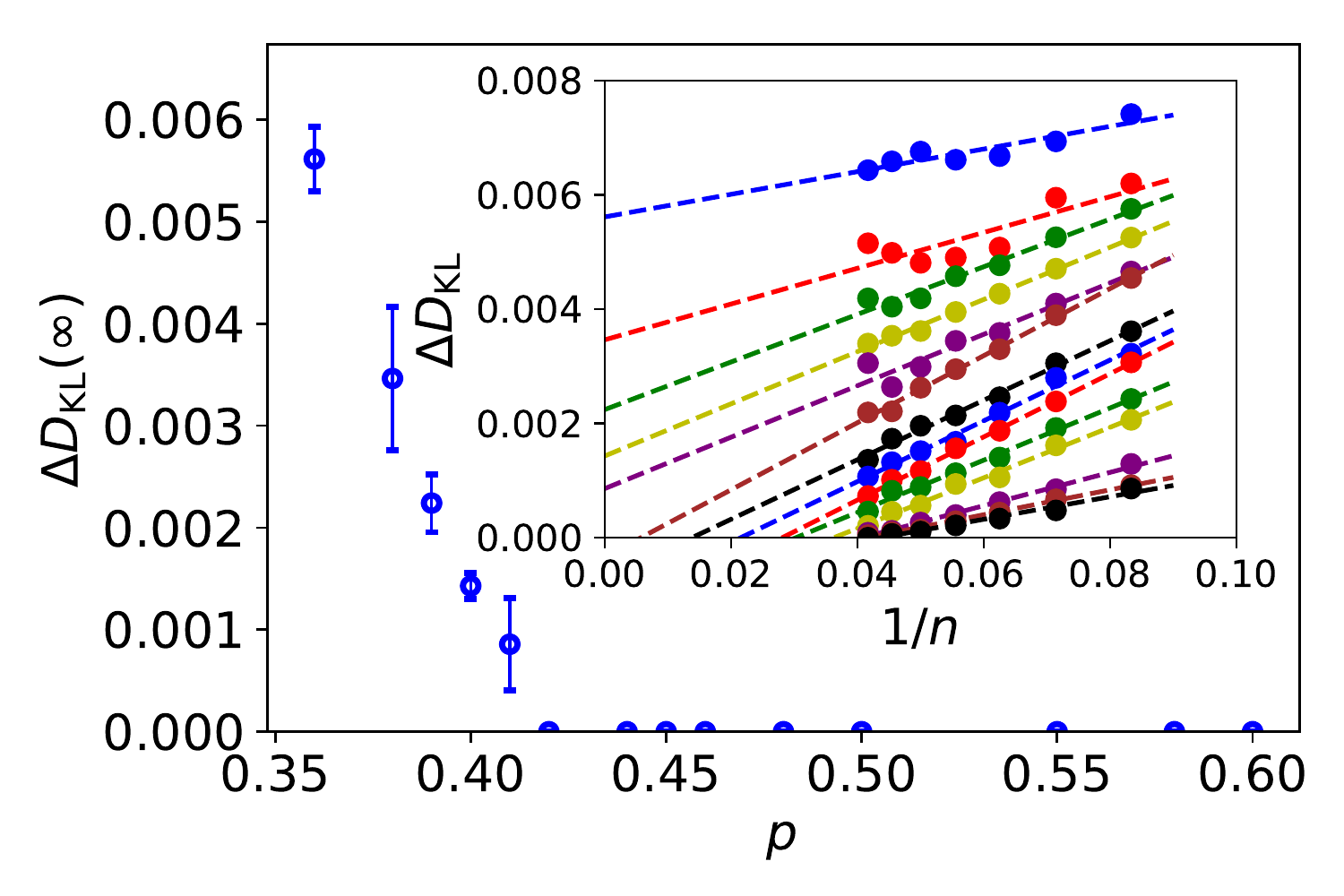}
}
\subfloat[][]{
\includegraphics[width=.33\textwidth]{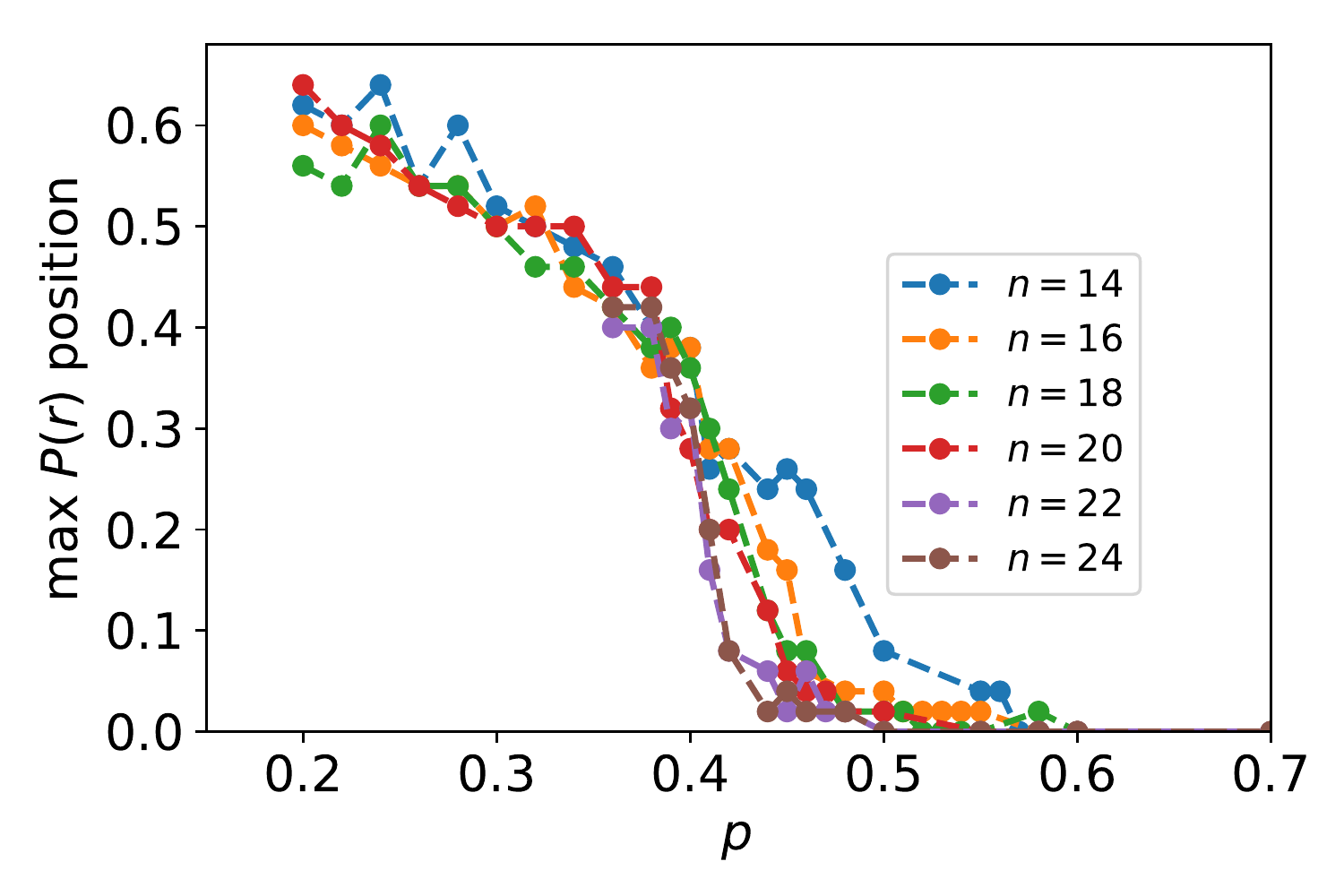}
}
\caption{(a) KL divergence $\Delta D_{KL}$ between $P(r)$ of evolved
  states and that of Poisson distribution as a function of projection
  rate $p$ at different system size.  (b) $\Delta D_{KL}$ at infinite
  size as a function of projection rate $p$. Inset: finite-size
  scaling for $\Delta D_{KL}$. (c) Position of maximum $P(r)$ as a
  function of projection rate $p$. The position is obtained from a
  histogram plot (not shown).
%Inset: $p_{0}$, the smallest projection rate that $P(r)$ reaches 0, as a function of $1/n$.
%{\color{red} More data are running to get a more precise $p_0$ value.}
}
\label{fig:DKL}
\end{figure*}
%%%%%%%%%%%%%%%%%%%%%%%%%%%%%%%%%%%%%%%%%%%%%%%%%%%%%%%%%%%%%%%%%%%%%%%%%%

The primary result in our work is the discovery of an intermediate
regime between $p_S < p < p_c$, where the ESS smoothly transitions
from the GUE distribution to the Poisson. We call this phase the
\emph{residual repulsion phase}. As shown in
Fig.~\ref{fig:spectrum}b,e, the strict level repulsion emblematic of
the chaotic regime disappears, i.e. $P(r)$ becomes nonzero for
$r\rightarrow 0$, and is replaced by a distribution that is between
the two regimes, having a maxima at a non-zero value of $r$. For the
quantum circuits taken from the set $\mathbf{U}_{\rm{syc}}$, this
transitionary phase exists within a shifted window of $p$ values, specifically $0.15 < p < 0.3$. We
infer that this quicker transition occurs as a result of the
particular universal gate set which we have selected. The argument is as
follows: from $\mathbf{U}_{\rm{syc}}$, the gate responsible for
introducing entanglement into the system is the $\rm{fSim}$ gate,
which is an $i\rm{SWAP}$ gate concatenated with a controlled $Z$,
having an internal block structure as $1\times 1,2\times 2,1\times
1$. The entanglement created by this gate is not as robust as that
introduced by a random Haar unitary gate, having no predefined
symmetries or structure. The entanglement arising from this structured
gate is therefore more susceptible to the presence of projective
measurements. Similar evidence for this argument is also found when considering a separate universal set $\mathbf{U} = \rm{CNOT},\rm{T},\rm{Hadamard}$. For this gate set, the
entangling gate ($\rm{CNOT}$) has an internal structure of $2\times 2,
2\times 2$, and only encodes a bit flip operation. In this case, we found that 
the residual repulsion phase exists only within the narrow window
$0.01 < p < 0.03$. In light of these narrower transition windows for the gate sets $\mathbf{U}_{\rm{syc}}$ and $\mathbf{U}$, we choose to focus our attentions on the Haar-random circuits when investigating the thermodynamic limit of the ES.

The Kullback-Leibler (KL) divergence, defined as
\begin{equation}
D_{KL}(P(x)||Q(x)) = \sum_x
P(x)\mathrm{ln}\left(\frac{P(x)}{Q(x)}\right) \,,
\end{equation}
provides a measure of distance between two distributions $P(x)$ and
$Q(x)$. If we calculate the $D_{KL}$ between numerically calculated
ESS distributions in the residual repulsion phase and the Poisson
distribution, there should exist a point $p$ where $D_{KL}$ goes to
zero, indicating that the numerical distributions have definitively
become Poisson. However, as previously mentioned, the numeric
distributions will necessarily deviate from the Poisson distribution
for small-$n$ circuits, such that the $D_{KL}$ takes on a nonzero even
in the Poisson phase. We therefore instead calculate the quantity
\begin{equation}
\Delta D_{KL} =D_{KL}(P_{\mathrm{final}}||P_{\mathrm{Poisson}}) -
\\ D_{KL}(P_{\mathrm{2layer}}||P_{\mathrm{Poisson}}) \,,
\end{equation}
where $P_{\mathrm{final}}$ is the calculated final state ESS
distribution and $P_{\mathrm{2layer}}$ is the ESS distribution
obtained by evolving random initial $n$-qubit product states with a
single time step of Haar-random two-qubit gates. This quantity is
positive in the residual repulsion phase but vanishes in the Poisson
phase, and can hence be used as an indicator of the transition between
the two phases with varying $p$.

In Fig.~\ref{fig:DKL} we locate the transition point $p_c$ by use of
two distinct figures of merit. The first one is the aforementioned
$\Delta D_{KL}$, shown in Fig.~\ref{fig:DKL}a. In the inset of
Fig.~\ref{fig:DKL}b, we plot $\Delta D_{KL}$ as a function of $1/n$
and use the data for various $n$
%to reveal finite size effect. The projection rate $p$ is ranged
%between $0.36$ and $0.6$. As shown by straight fitting lines, the
%data approximated fall on straight lines with similar slope at large
%$\Delta D_{KL}$, for example $p<0.5$. When $\Delta D_{KL}$ is close
%enough to 0, for instance $p>0.55$, the relation $\Delta D_{KL} \sim
%k/n$ tends to bend flat with small slope $k$. $\Delta D_{KL}$ at
%infinite size $n\rightarrow \infty$ can be obtained from
to extrapolate linearly to $n\to\infty$. For small $p$, $\Delta
D_{KL}$ extrapolates to a positive value at infinite size.
%If the intersection is negative, as $\Delta D_{KL}$ can not be
%negative to get physical meaning, we propose that negative
%intersection indicates $\Delta D_{KL}$ converges to 0 at infinite
%size. The slope of relation $\Delta D_{KL} \sim k/n$ will Ruce when
%$\Delta D_{KL} $ is close to zero, as shown in numerical results at
%large $p$. Based on this argument, the results in Fig.~\ref{fig:DKL}
%(b) indicate a phase transition at $p\approx 0.41$.
The slope of the finite-size scaling curve increases with increasing
$p$ and at $p_c \simeq 0.41$ the curves start intersecting the $\Delta
D_{KL}=0$ axis at finite $n$, indicating a transition to the Poisson
phase. Additionally, in Fig.~\ref{fig:DKL}c we show how the position
of the maximum of $P_{\mathrm{final}}(r)$ changes with $p$.
%We use this position as  a parameter to quantify how distribution $P(r)$ changes.
For the Poisson distribution the maximum is at $r=0$, whereas the GUE
distribution has its maximum $P(r)$ at $r =(\sqrt5-1)/2\approx
0.618$. Fig.~\ref{fig:DKL}c shows that the maximum of
$P_{\mathrm{final}}(r)$ decreases from $r\approx0.618$, reaching zero
close to $p_c$.
%To get the finite size effect, we plot $p_0$, the minimum projection
%rate required to get peak of $P(r)$ at 0, as a function of inverse
%system size $1/n$. {\color{red} At infinite size, $p_0$ gets to 0.4,
%which is another evidence of phase transition at $p_c \approx 0.4$}

%%%%%%%%%%%%%%%%%%%%%%%%%%%%%%%%%%%%%%%%%%%%%%%%%%%%%%%%%%%%

\section{ESS transition at $p_c$ and bond percolation in spacetime}
\label{sec:discussion} 

We propose that the transition at $p_c$ can be explained as bond
percolation in a square lattice~\cite{Kesten1980}. Percolation theory,
however, predicts a transition at $p=0.5$ and not at the observed $p_c
\approx 0.41$. The reason for this discrepancy is that a single
projector may affect multiple bonds.

%%%%%%%%%%%%%%%%%%%%%%%%%%%%%%%%%%%%%%%%%%%%%%%%%%%%%%%%%%%%%%%%%%%%%%%%%5
\begin{figure}[tb]
\centering
\subfloat[][]{
\includegraphics[width=0.9\columnwidth]{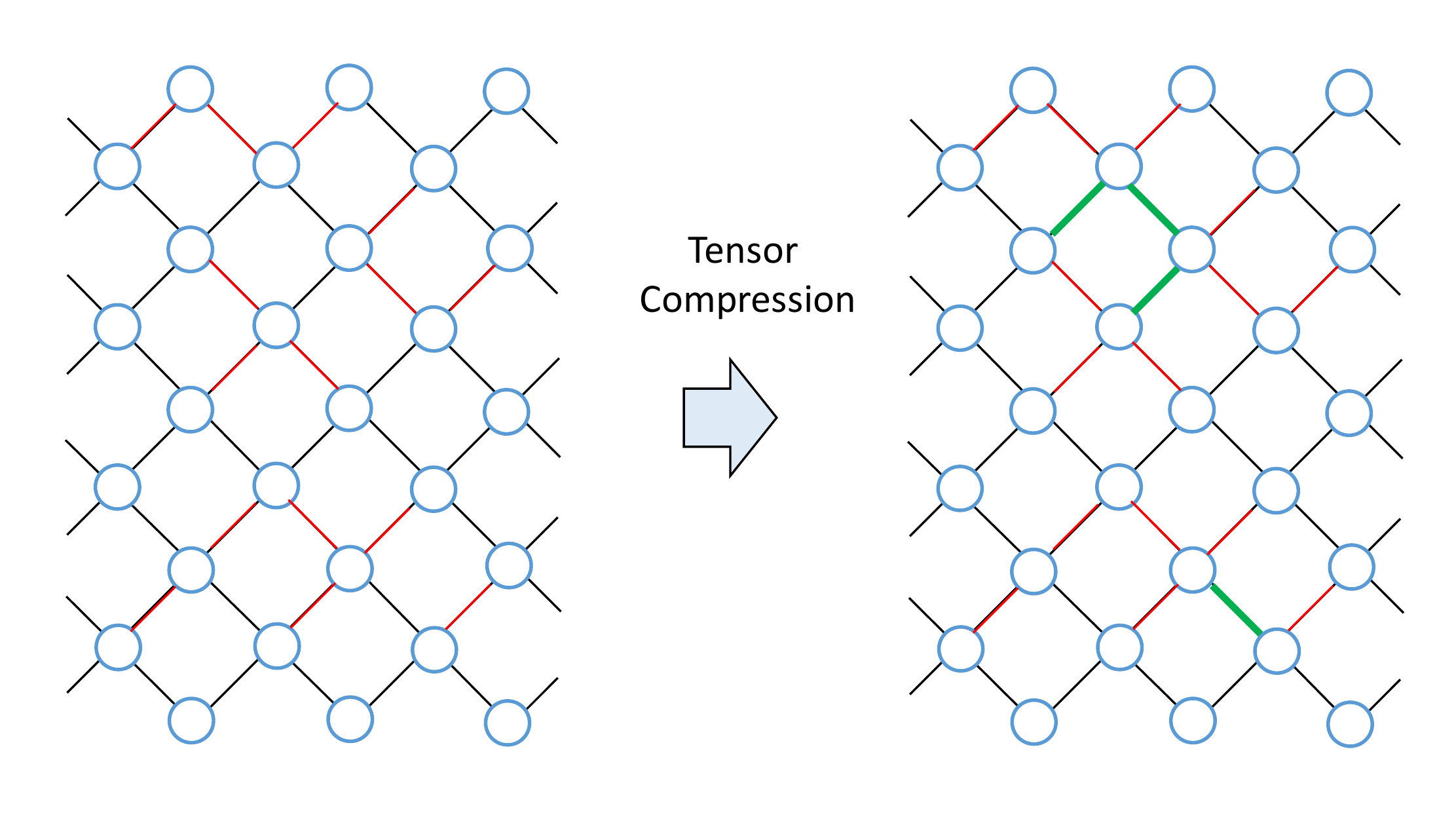}
}\\
\subfloat[][]{
\includegraphics[width=0.9\columnwidth]{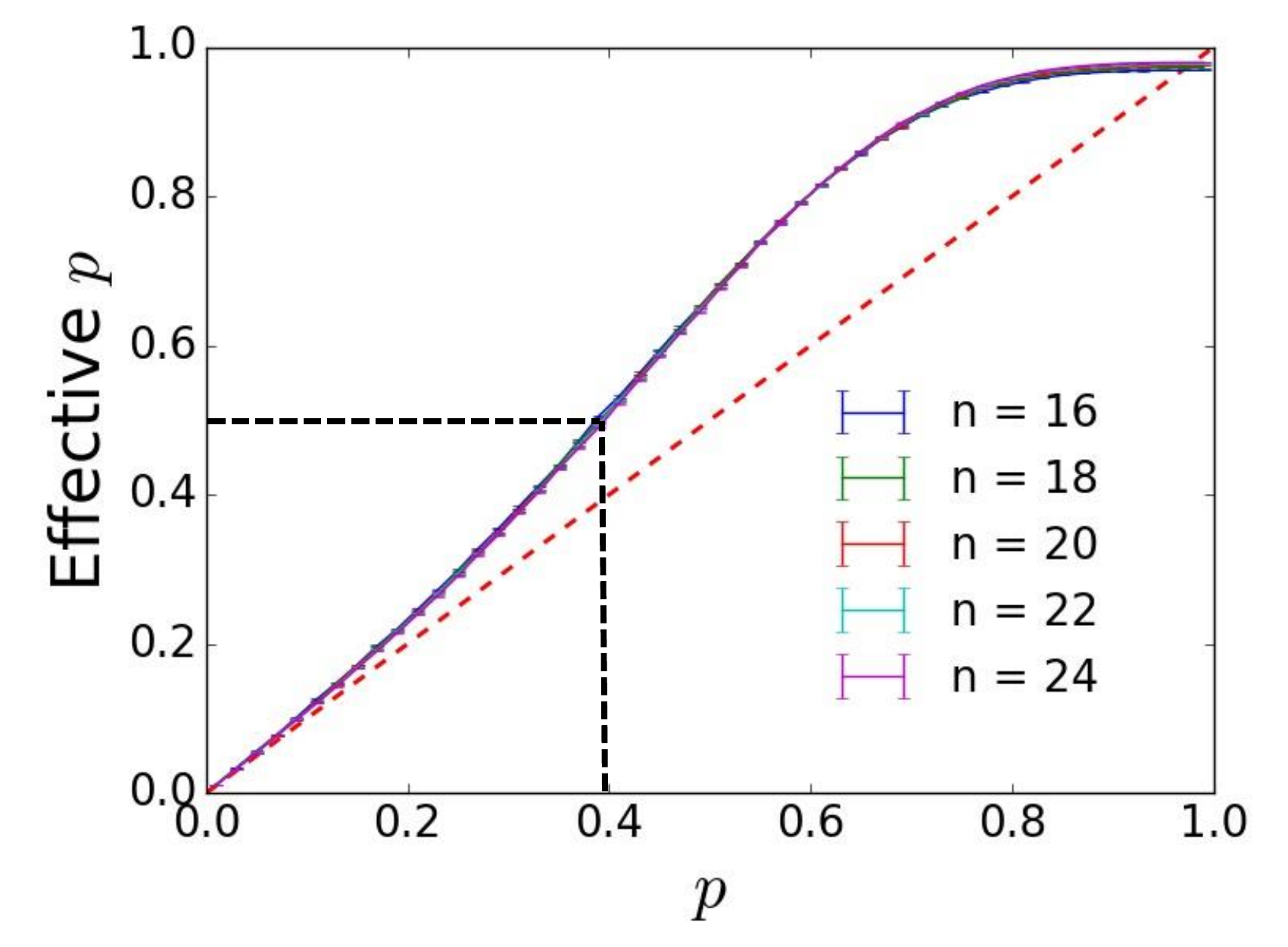}
}
\caption{(a) Bond dimension distribution before (left) and after
  (right) compression in a tensor network corresponding to circuit
  with projectors inserted in bonds marked red. Bonds whose dimension
  is reduced to 1 after compression are indicated by green lines. (b)
  Numerical results for effective projection rate as a function of
  projector insertion rate $p$.
%Effective projection rate is defined as ratio of dimension 1 bond
%after cleaning.  Dashed blue lines indicate that $p$ only needs to be
%around $0.4$ to get a 0.5 effective ratio.
}
\label{fig:cleaning}
\end{figure}
%%%%%%%%%%%%%%%%%%%%%%%%%%%%%%%%%%%%%%%%%%%%%%%%%%%%%%%%%%%%%%%%%%%%%%%%%5

%In Fig.~\ref{fig:mapping} of mapping circuits to tensor network, we
%split tensor $T_{g}$ into two tensors and add additional bond to make
%tensor network in regular square lattice form.

To illustrate this, we consider the more direct mapping from circuit
to tensor network shown in Fig.~\ref{fig:cleaning}a, which implements
only the first step of Fig.~\ref{fig:mapping}a. Without projectors,
this yields a rotated square lattice with uniform bond dimension
2. Adding a projection operator to a bond reduces the dimension of
that bond to 1. Naively, one would expect that percolation occurs when
a giant component of bonds with a projector forms, which for the
square lattice would happen at density $p=0.5$. However, the
projection of a qubit to the computational basis has a disentangling
effect also in its vicinity in spacetime and not only at the
particular point of insertion of a projector. This effect is resolved
by the compression step of our algorithm. In the example of
Fig.~\ref{fig:cleaning}a, after the compression step, all bonds
indicated by a green line are also reduced to dimension 1. This
reasoning suggests a modified percolation threshold based on the
density of dimension-1 bonds after a compression sweep, which we call
the \emph{effective} projection rate.

We verify this numerically. Fig.~\ref{fig:cleaning}b shows the
effective projection rate as a function of the density of projectors
$p$.
%Effective projection rate is defined as ratio of dimension 1 bond
%after cleaning.  We propose that effective projection rate is the one
%directly related to 2D bond percolation transition. For 2D square
%lattice, percolation theory predicts a phase transition at bond
%connection ratio of 0.5.
Dashed black lines indicate that an effective projection ratio of 0.5
is achieved at $p \approx p_c$. This agrees well with our ESS-based
estimate for the transition point. As the two-dimensional bond
percolation transition is characterized by absence of long range
correlations, this reflects the fact that the system becomes integrable at
$p>p_{c}$. It should be noted that the compression algorithm is exact
to machine precision and hence accuracy does not factor appreciably
into this reasoning.
%, and thus our computational results are independent on methods one
%use.

%From the aspect of ESS, the Poisson phase at $p>p_{c}$ shows similar
%features to Anderson localization. As researched in \cite{ZCY2017},
%quantum states evolved by Anderson localization Hamiltonian show
%integrable feature with Poisson distribution in ESS.  It agrees well
%with the simple argument that quantum system becomes localized when
%projection is strong enough. However the residual repulsion phase at
%$p_S<p<p_{c}$ has ESS features different from other known
%localization. For the case of many-body localization, states quenched
%by many-body localization Hamiltonian \cite{ZCY2017} show
%Wigner-Dyson distribution, while eigenstates of the Hamiltonian
%\cite{Geraedts2016} have semi-Poisson ESS. Both these two many-body
%localization related cases have strict level repulsion at
%$r\rightarrow 0$.

%%%%%%%%%%%%%%%%%%%%%%%%%%%%%%%%%%%%%%%%%%%%%%%%%%%%%%%%%%%%%%%%%%
\section{conclusion}\label{sec:conclusion}

The work presented here explores projection-driven quantum circuits
from the perspective of the ESS of the output state. Our results
uncover three distinct behaviors of the entanglement spectrum with
increasing the rate, $p$, of projection of qubits to the computational
basis. The first regime, $0 < p < p_S$, displays volume-law
entanglement entropy and Wigner-Dyson statistics of the EES. At
$p=p_S$ the systems undergoes a volume-to-area-law transition similar
to that studied in Refs.~\cite{Cao2018, Chan2018, Skinner2018, Li2019,
  Li2018, Bao2019, Jian2019, Gullans2019}. The principal result of
this paper is that the ESS in the area law phase emerging at $p=p_S$
is non-universal and interpolates between Wigner-Dyson and Poisson
statistics, with the region of residual level repulsion extending up
to a second transition, $p=p_c > p_S$, beyond which the ESS of the
system is Poisson.
%That there is such a transition is intuitively anticipated by
%considering the limit of infinite projection rate, in which case
%there is no entanglement in the output due to a quantum Zeno effect.

This second transition within the area law
  phase is revealed by repartioning the system randomly into two subsystems and
  probing the entanglement level statistics. In particular,
our tensor network algorithm, which resolves entanglement by
monitoring the distribution of bond dimensions across the 1+1D
spacetime of the circuit, allows us to interpret this transition
% between non-universal and Poisson statistics 
as a percolation transition of
entangled bonds and to locate the corresponding critical value $p=p_c$
via finite size scaling.
We note that, unlike previous results in
Refs.~\cite{Bao2019,Jian2019} obtained in the limit of large local
Hilbert space dimension that associate the volume-to-area law
transition to percolation in a classical Potts model, here we find
that the ESS transition from non-universal to Poisson statistics is
due to percolation of entangled bonds in the circuit spacetime itself.

This work leaves open the question of the origin of the intermediate
regime with non-universal statistics of the entanglement spectrum. The
nature of level statistics is determined by the details of the
interactions between eigenvalues which can induce complex
non-universal level statistics, including a Griffiths-like phase
observed in studies of MBL \cite{ Demler2016, Serbyn2016}. More detailed work is
needed to elucidate the non-universal regime in the ESS identified in
this paper.
Nevertheless, the methodology and findings presented here enable a more fine-grained characterization of entanglement buildup in noisy quantum circuits compared to analyses based solely on entanglement entropy, and can be used to study near-term noisy quantum chips via quantum state tomography.

%%%%%%%%%%%%%%%%%%%%%%%%%%%%%%%%%%%%%%%%%%%%%%%%%%%%%%%%%%%%%%%%%%
\acknowledgments

We thank David Huse for constructive comments on the manuscript
  that helped identify where  clarification was needed. J.A.R and
E.R.M. acknowledge partial financial support from NSF grant
No. CCF-1844434. L.Z., C.C., and A.E.R acknowledge partial financial
support from NSF grant No. CCF-1844190. Numerical calculations were
performed on the Boston University Shared Computing Cluster, which is
administered by Boston University Research Computing Services.

%%%%%%%%%%%%%%%%%%%%%%%%%%%%%%%%%%%%%%%%%%%%%%%%%%%%%%%%%%%%%%%%%%
%\bibliography{biblio}


%%%%%%%%%%%%%%%%%%%%%%%%%%%%%%%%%%%%%%%%%%%%%%%%%%%%%%%%%%%%%%%%%%

\end{document}